\documentclass{article}
\usepackage{spconf,amsmath,graphicx}
\usepackage{multirow}
\usepackage{array}
\usepackage{url}
\usepackage{amsmath}
\usepackage{color}
\usepackage{booktabs}
\usepackage{hyperref}

\newcolumntype{P}[1]{>{\centering\arraybackslash}p{#1}}

\title{Enhancing Low-Quality Voice Recordings Using Disentangled Channel Factor and Neural Waveform Model}
\name{Haoyu Li$^1{^,}^2$, Yang Ai$^3$, Junichi Yamagishi$^1{^,}^2$}
\address{$^1$National Institute of Informatics, Japan
$^2$SOKENDAI, Japan\\
$^3$University of Science and Technology of China, P.R.China}
\begin{document}
%
\maketitle
\begin{abstract}
High-quality speech corpora are essential foundations for most speech applications. However, such speech data are expensive and limited since they are collected in professional recording environments.
In this work, we propose an encoder-decoder neural network to automatically enhance low-quality recordings to professional high-quality recordings. To address channel variability, we first filter out the channel characteristics from the original input audio using the encoder network with adversarial training. Next, we disentangle the channel factor from a reference audio. Conditioned on this factor, an auto-regressive decoder is then used to predict the target-environment Mel spectrogram. Finally, we apply a neural vocoder to synthesize the speech waveform. Experimental results show that the proposed system can generate a professional high-quality speech waveform when setting high-quality audio as the reference. It also improves speech enhancement performance compared with several state-of-the-art baseline systems. 
\end{abstract}
\vspace{-0.75mm}
\section{Introduction}
\label{sec:intro}
\vspace{-1mm}
Recently, deep neural networks (DNNs) have been widely used in many speech-generation tasks, such as text-to-speech synthesis \cite{ze2013statistical} and voice conversion \cite{chen2014voice}. Such a deep-learning based data-driven approach typically requires a large amount of high-quality expensive speech data for training. 
Although self-recorded or public found speech can be relatively easily obtained, it is often recorded in uncontrolled environments by using non-professional microphones, where acoustic noise, room reverberation, and bad frequency characteristics of the recording device inevitably degrade audio quality. 

To enhance low-quality recordings, traditional speech enhancement methods, e.g., weighted linear prediction error (WPE) \cite{nakatani2010speech}, spectral subtraction \cite{boll1979suppression}, and log-MMSE speech magnitude estimator \cite{ephraim1985speech}, have been extensively studied. These signal-processing methods are typically developed for a single application scenario, such as speech denoising (e.g. \cite{boll1979suppression,ephraim1985speech}), de-reverberation (e.g. \cite{nakatani2010speech}), or audio effect adaptation (e.g. equalization \cite{verfaille2006adaptive}). Although one can combine denoising, de-reverberation, and equalization methods to sequentially address each subproblem, Mysore \MakeLowercase{\textit{et al.}} \cite{mysore2014can} pointed out that such intuitive combination would degrade audio quality due to undesired synergy between processes. For example, a sound equalizer might amplify background noise by wrongly amplifying noisy-frequency components, which causes conflict with the speech denoising process.
The performance of such traditional methods is still far from satisfactory.

Our goal for this work is to automatically enhance low-quality recordings by simultaneously removing noise and reverberation and applying pleasing audio effects. More specifically, we propose an encoder-decoder neural network to directly transform the input recordings to sound as if they were produced under other recording conditions (e.g. in a professional studio). Recording conditions, including noise, reverberation, microphone characteristics, and audio effects, are jointly considered, which we collectively refer to as the \textit{channel factor}. The target recording condition is derived from an additional reference audio and can be represented as the channel factor (embedding) using a channel modeling (CM) network. In the inference stage, the input recording is first filtered out its original channel characteristics by an adversarially trained encoder, and then passed to the decoder to predict the target-environment Mel spectrogram, conditioned on the channel factor obtained from the reference. Finally, we use a \mbox{WaveRNN} vocoder to generate a speech waveform from the predicted Mel spectrogram. With this flexible framework, we can not only enhance low-quality audio by providing high-quality clean audio as the reference but also transfer the audio effect, e.g., adapting the input recording to the new reverb effect if we designate the appropriate reference audio.

This paper is organized as follows: Section~\ref{sec:rw} reviews the relevant work. Section~\ref{sec:overview} describes the proposed system and gives details of each network component. Section~\ref{sec:exp} presents the experimental results. We conclude our work and discuss future direction in Section~\ref{sec:conclu}.

\vspace{-2.5mm}
\section{Related work}
\label{sec:rw}
\vspace{-1mm}
Many DNN-based speech enhancement methods \cite{xu2014regression,weninger2015speech} have recently been proposed for directly predicting clean speech representations from noisy input and significantly outperform traditional methods.
Nevertheless, most of these methods operate on the magnitude spectrogram but disregard the phase. Consequently, noisy phase distortion is inevitably introduced by inverse short-time Fourier transform (ISTFT), which degrades performance. To overcome such limitations, end-to-end waveform models, such as SEGAN \cite{pascual2017segan,sarfjoo2019transformation} and Wavenet \cite{oord2016wavenet}, have attracted significant attention. Very recently, Su \MakeLowercase{\textit{et al.}} \cite{su2019perceptually, su2020hifi} proposed a Wavenet-based waveform-to-waveform mapping system for speech enhancement and achieved good results. Compared with time-frequency features, however, the raw waveform samples are redundant \cite{yin2020phasen}, therefore difficult to model, and the trained system tends to be more susceptible to overfitting.

Different from previous studies \cite{sarfjoo2019transformation, su2019perceptually, su2020hifi}, we choose the Mel spectrogram on which to operate since this high-level representation is much smoother than raw samples and easier to handle. To alleviate the phase distortion introduced by ISTFT, we use the state-of-the-art \mbox{WaveRNN} vocoder \cite{kalchbrenner2018efficient} to synthesize the waveform. Maiti \MakeLowercase{\textit{et al.}} \cite{maiti2020speaker} also proposed applying neural vocoders for speech enhancement. Unlike their work, we do not focus on investigating the effect of different neural vocoders in the waveform-synthesis stage but on studying the model architecture and training objective in the spectrogram-enhancement stage. For waveform synthesis, we simply select WaveRNN as the vocoder.

\vspace{-1mm}
\section{System Overview}
\label{sec:overview}

\begin{figure}[t]

\begin{minipage}[b]{1.0\linewidth}
  \centering
  \centerline{\includegraphics[width=1.0\linewidth]{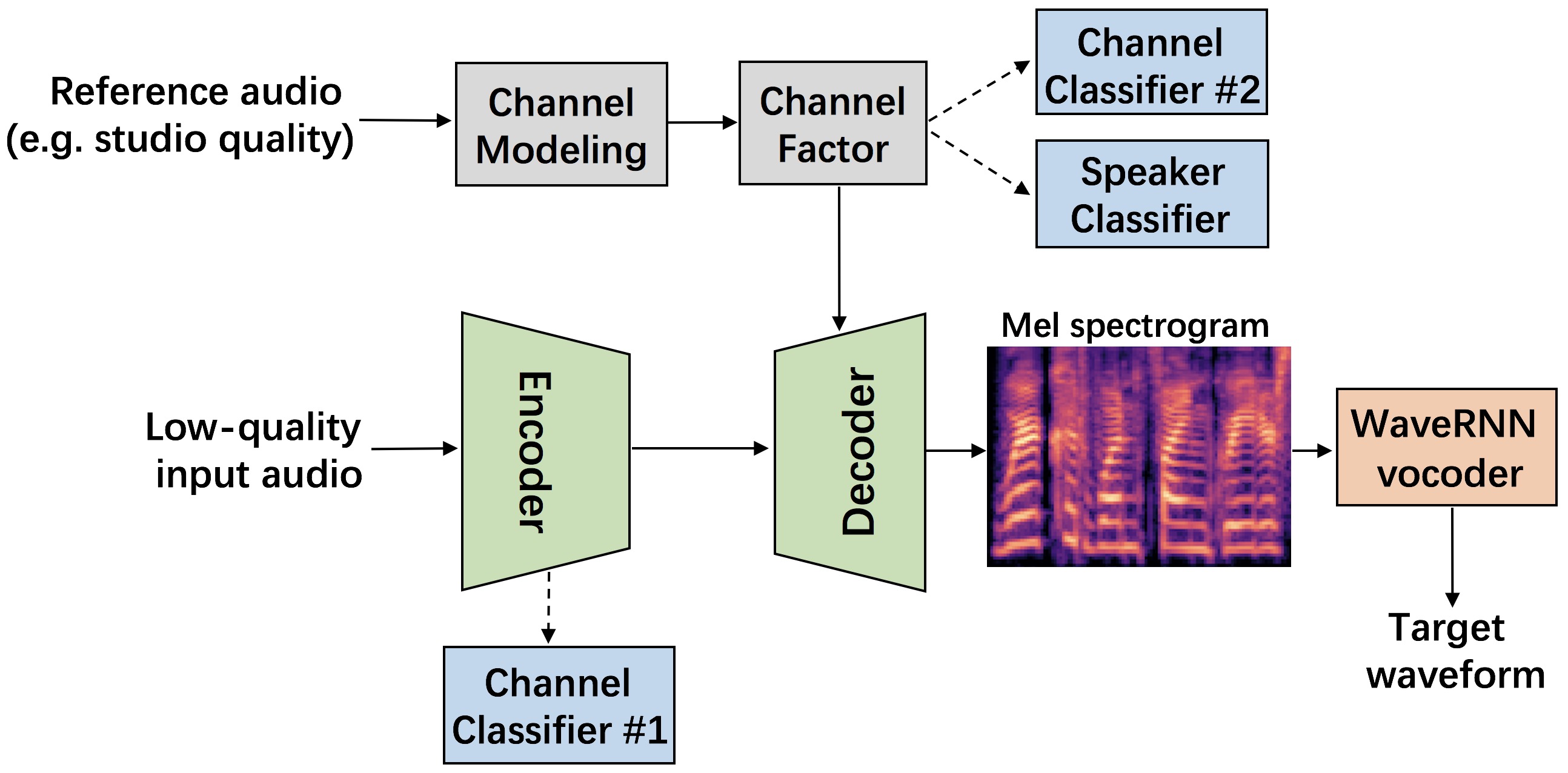}}
\end{minipage}

\caption{Overall diagram of proposed system.}
\label{fig:diagram}
\vspace{-3mm}
\end{figure}

The system diagram is illustrated in Fig.~\ref{fig:diagram}. It consists of three main components: an encoder, a channel modeling (CM) network, and a decoder. In addition, a WaveRNN vocoder works separately as the waveform synthesis module. 
\vspace{-1mm}
\subsection{Encoder}
\label{sec:encoder}

The encoder is designed to filter out the channel characteristics from input audio. More specifically, the input audio is first transformed to the magnitude spectrogram by STFT, and then passed to the encoder to produce the channel-invariant features. The encoder consists of a six-layer 2D CNN, each layer with batch normalization, ReLU activation, and zero paddings, and one bi-directional LSTM (Bi-LSTM) layer. The details of the encoder's parameters are listed in Table~\ref{tab:param-enc}. 

To encourage the encoder to produce channel-invariant features, inspired by \cite{liao2018noise} and \cite{hsu2019disentangling}, we introduce a \textbf{channel classifier (\#1)} as a discriminator for adversarial training. It consists of one uni-directional LSTM (Uni-LSTM) layer with 400 nodes and one fully-connected layer with a softmax layer, which predicts the channel type (recording condition) of the input audio. In the training stage, this classifier is optimized to accurately predict the channel type by minimizing the cross-entropy classification loss. On the other hand, the encoder is optimized oppositely to maximize the classification loss to prevent the produced features from encoding channel information. This adversarial training encourages the encoder to filter out the channel information from its input.

\begin{table}[t]
    \caption{Parameters of encoder. Kernel shape of 2D CNN layers is represented as [kernel size tuple, stride tuple, output channels]. $T$ and $F$ denote the number of frames and frequency bins, respectively.}
    \vspace{1mm}
    \label{tab:param-enc}
    \centering
    \renewcommand\arraystretch{1.25}
    \scalebox{0.83}{
    \begin{tabular}{c|c|c|c}
        \toprule
          \textbf{Layer} &
           \textbf{Input shape} &
           \textbf{Kernel shape / Nodes} &
           \textbf{Output shape}  \\

          \hline
          CNN 1 & (T, F) & [(1, 7), (1, 1), 64] & (64, T, F)  \\
          CNN 2 & (64, T, F) & [(7, 1), (1, 1), 64] & (64, T, F) \\
          CNN 3 & (64, T, F) & [(5, 5), (1, 1), 64] & (64, T, F) \\ 
          CNN 4 & (64, T, F) & [(5, 5), (1, 2), 64] & (64, T, F // 2) \\
          CNN 5 & (64, T, F // 2) & [(5, 5), (1, 2), 64] &  (64, T, F // 4) \\
          CNN 6 & (64, T, F // 4) & [(1, 1), (1, 1), 8] & (8, T, F // 4) \\
          Bi-LSTM & (T, F $\times$ 2) & 256 & (T, 512) \\
        \bottomrule
    \end{tabular}
    }
    \vspace{-1mm}
\end{table}
\vspace{-1mm}
\subsection{Channel Modeling}
\label{sec:channel-net}

\begin{figure}[t]
\begin{minipage}[b]{1.0\linewidth}
  \centering
  \centerline{\includegraphics[width=1.0\linewidth]{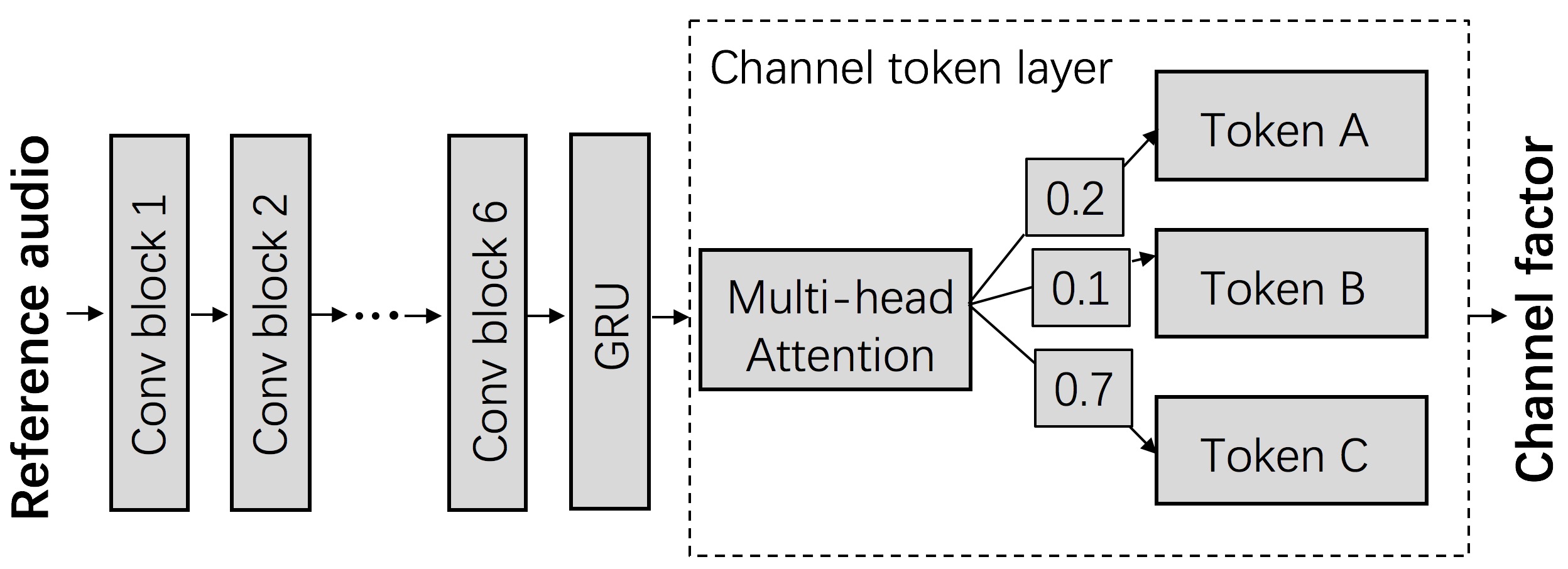}}
\end{minipage}
\caption{Detailed structure of CM network.}
\label{fig:channel-network}
\vspace{-2.7mm}
\end{figure}

The channel modeling (CM) network explicitly extracts the channel factor from the reference audio. Its structure is shown in Fig.~\ref{fig:channel-network}. 
We use this model structure since it effectively encodes latent factors \cite{wang2018style,li2020noise}.
Instead of using a one-hot code, the channel factor can be automatically encoded as a neural code from the reference audio, which enables the system to deal with the unseen channel condition and unlabelled reference audio. Moreover, the CM network can be jointly optimized with other neural components, which further provides better results.

The network takes as input the magnitude spectrogram computed from the reference. It consists of a six-layer 2D CNN each with a 5$\times$5 kernel, 2$\times$2 stride, batch normalization, and ReLU activation. The output channels are set to 32, 32, 64, 64, 128, and 128, respectively. A uni-directional gated recurrent unit (GRU) layer with 128 nodes follows the last CNN layer, producing an intermediate feature. Next, a channel token layer is added, which consists of 12 trainable channel tokens and a multi-head attention module \cite{vaswani2017attention}. Specifically, each token has 256 dimensions, and the number of attention heads is set to 8. The intermediate feature output by the GRU layer is fed to the channel token layer and serves as the \textit{query} vector, then the attention module calculates the similarity (weight) between the query and each token. Finally, the channel factor (vector) is formed as the weighted sum of these channel tokens. 

To better disentangle channel and speaker identities from the reference audio, we introduce two additional classifiers, \textbf{channel classifier (\#2)} and \textbf{speaker classifier}. Both are fully-connected networks with one 256-node hidden layer followed by a softmax layer to predict the channel type or speaker identity. Note that different from channel classifier (\#1) used in the encoder, classifier (\#2) here encourages the channel factor to be more informative about channel information. While speaker classifier still serves as the adversarial discriminator to filter out the speaker information from the channel factor.
\vspace{-2.00mm}

\subsection{Decoder}
\label{sec:decoder}
\vspace{-0.85mm}

\begin{figure}[t]
\begin{minipage}[b]{1.0\linewidth}
  \centering
  \centerline{\includegraphics[width=0.95\linewidth]{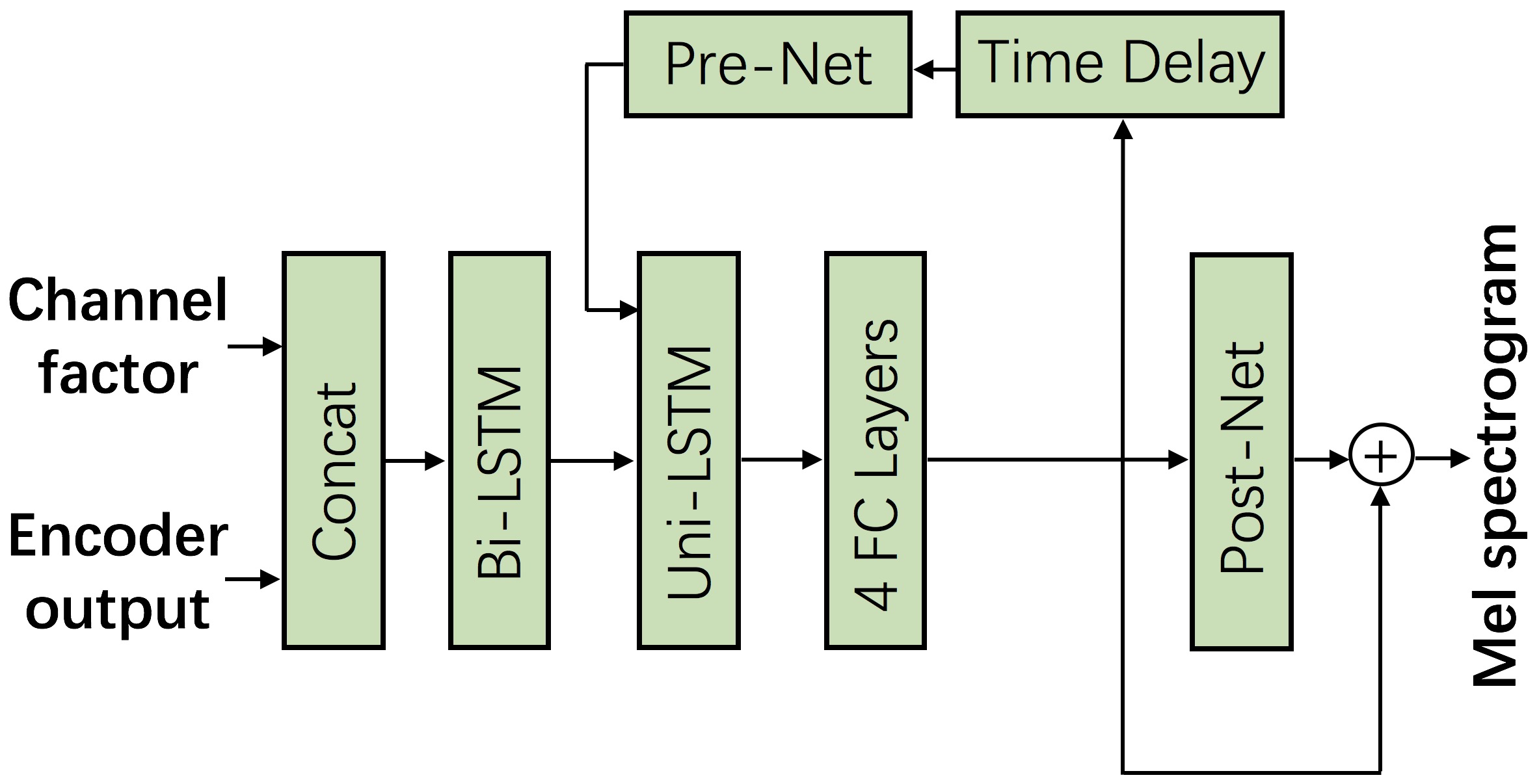}}
\end{minipage}
\caption{Decoder structure. Concat and FC denote concatenation operation and fully-connected layer, respectively.}
\label{fig:decoder-network}
\vspace{-3mm}
\end{figure}

The auto-regressive decoder shown in Fig.~\ref{fig:decoder-network} is used to produce the target-environment Mel spectrogram, which shares similar channel characteristics to those of the reference audio. The extracted channel factor is first repeatedly concatenated to the encoder output in every time frame. The resulting concatenated features are processed by a Bi-LSTM layer with 256 nodes, and then passed to a Uni-LSTM layer with 512 nodes. Four fully-connected layers are sequentially added, each with 80 nodes, to produce the 80-dimensional Mel spectrogram. Similar to Tacotron2 \cite{shen2018natural}, we add a 2-layer Pre-Net each with 256 nodes for the auto-regressive process. The produced Mel spectrogram from the previous time step is processed through Pre-Net and fed into the Uni-LSTM layer for the prediction of the current time step. A five-layer convolutional Post-Net module used in Tacotron2 is also introduced to predict the Mel spectrogram residual to improve the overall reconstruction. 
\vspace{-1.25mm}

\vspace{-1mm}
\subsection{WaveRNN Vocoder}
\label{sec:wavernn}
\vspace{-0.8mm}
To avoid introducing the noisy phase, a WaveRNN vocoder is used to directly generate the waveform from the Mel spectrogram. We use a speaker-independent WaveRNN, which effectively generalizes to unseen speakers. 
Since this vocoder is trained using a large external corpus, compared to the deterministic ISTFT, such a robust neural waveform model has better tolerance to the prediction error of spectrogram, and thus is able to generate a higher quality waveform.

\subsection{Training Objective}
\label{sec:train-obj}

To state the training objective of our proposed system, we review each component shown in Fig.~\ref{fig:diagram} and use the following definitions:
\begin{align}
 \boldsymbol{z}_e^{1:T_i}=\boldsymbol{\Phi}_e(\boldsymbol{o}_{in}^{1:T_i}),\\
 \boldsymbol{z}_c=\boldsymbol{\Phi}_c(\boldsymbol{o}_{ref}^{1:T_r}),\\
 \boldsymbol{\hat{m}}^{1:T_i}=\boldsymbol{\Phi}_d(\boldsymbol{z}_e^{1:T_i},\boldsymbol{z}_c)
\end{align}
The encoder $\boldsymbol{\Phi}_e$ encodes the input spectrogram of $T_i$ frame length, $\boldsymbol{o}_{in}^{1:T_i}=\{\boldsymbol{o}_{in}^1, \cdots, \boldsymbol{o}_{in}^{T_i}\}$, into a latent sequence $\boldsymbol{z}_e^{1:T_i}=\{\boldsymbol{z}_{e}^1, \cdots, \boldsymbol{z}_{e}^{T_i}\}$.
The CM network $\boldsymbol{\Phi}_c$ extracts the channel factor (vector) $\boldsymbol{z}_c$ from the reference spectrogram of $T_r$ frame length, $\boldsymbol{o}_{ref}^{1:T_r}=\{\boldsymbol{o}_{ref}^1, \cdots, \boldsymbol{o}_{ref}^{T_r}\}$. 
Decoder $\boldsymbol{\Phi}_d$ takes as inputs $\boldsymbol{z}_e^{1:T_i}$ and $\boldsymbol{z}_c$ and predicts the Mel spectrogram $\boldsymbol{\hat{m}}^{1:T_i}=\{\boldsymbol{\hat{m}}^1, \cdots, \boldsymbol{\hat{m}}^{T_i}\}$. 
We jointly optimize three modules, $\boldsymbol{\Phi}_e$, $\boldsymbol{\Phi}_c$, and $\boldsymbol{\Phi}_d$, to minimize the mean square error (MSE) between the predicted Mel spectrogram $\boldsymbol{\hat{m}}^{1:T_i}$ and ground truth $\boldsymbol{m}^{1:T_i}$, as formulated in Eq.\ \ref{eq:mse}.
\begin{equation}
\begin{split}
    \mathcal{L}_{MSE}&(\boldsymbol{\Phi}_e, \boldsymbol{\Phi}_c, \boldsymbol{\Phi}_d) = \frac{1}{T_i} \sum_{j=1}^{T_i} \Vert\boldsymbol{\hat{m}}^{j}-\boldsymbol{m}^{j}\Vert_2^2
\end{split}
\label{eq:mse}
\end{equation}

In addition to the main MSE objective, we add the following three objectives:
\begin{align}
 \mathcal{L}_{enc\_ch}&(\boldsymbol{\Phi}_e, \boldsymbol{D}_{c1})=CE(
 \boldsymbol{D}_{c1}(\boldsymbol{z}_e^{1:T_i}), \boldsymbol{c}_{in}), 
   \label{eq:eq5}
\\
    \mathcal{L}_{cm\_ch}&(\boldsymbol{\Phi}_c, \boldsymbol{D}_{c2})=CE(
    \boldsymbol{D}_{c2}(\boldsymbol{z}_c), \boldsymbol{c}_{ref}),
 \label{eq:eq6} 
\\
 \mathcal{L}_{cm\_spk}&(\boldsymbol{\Phi}_c, \boldsymbol{D}_{s})=CE(
 \boldsymbol{D}_{s}(\boldsymbol{z}_c), \boldsymbol{s}_{ref})
 \label{eq:eq7}
\end{align}
where $CE$ denotes the cross-entropy loss, and $\boldsymbol{D}_{c1}$, $\boldsymbol{D}_{c2}$, and $\boldsymbol{D}_{s}$ denote channel classifier \#1, \#2, and the speaker classifier, respectively. The channel types of the input and the reference are represented as one-hot labels, i.e., $\boldsymbol{c}_{in}$ and $\boldsymbol{c}_{ref}$, and the speaker label of the reference is denoted as $\boldsymbol{s}_{ref}$. As explained in previous sections, $\mathcal{L}_{enc\_ch}$ is used as the adversarial training objective to filter out the channel information from the encoder output $\boldsymbol{z}_e^{1:T_i}$. We also use $\mathcal{L}_{cm\_ch}$ as an auxiliary objective and $\mathcal{L}_{cm\_spk}$ as an adversarial objective, to encourage the channel factor $\boldsymbol{z}_c$ to encode more channel information but less speaker information. 
In the training stage, the neural components (i.e. $\boldsymbol{\Phi}_e$, $\boldsymbol{\Phi}_c$, and $\boldsymbol{\Phi}_d$) and classifiers (i.e. $\boldsymbol{D}_{c1}$, $\boldsymbol{D}_{c2}$, and $\boldsymbol{D}_{s}$) are optimized alternatively. At one training step, we optimize three classifiers individually by minimizing their corresponding cross-entropy objectives, which are $\mathcal{L}_{enc\_ch}$, $\mathcal{L}_{cm\_ch}$, and $\mathcal{L}_{cm\_spk}$. At the next training step, we fix the classifiers and jointly optimize all three neural components with the following training objective:
\begin{equation}
    \mathcal{L}=\mathcal{L}_{MSE}+\alpha*\mathcal{L}_{cm\_ch}-\beta*\mathcal{L}_{enc\_ch}-\gamma*\mathcal{L}_{cm\_spk},
\label{eq:all-obj}
\end{equation}
where $\alpha$, $\beta$, and $\gamma$ are hyper-parameters controlling the weights of different sub-objectives.
\vspace{-0.78mm}
\section{Experiments}
\label{sec:exp}
\vspace{-2.0mm}
\subsection{Dataset}
\label{sec:dataset}
\vspace{-0.5mm}
The DAPS (device and produced speech) dataset \cite{mysore2014can} was used in our experiments. It provides aligned recordings of high-quality speech\footnote{High-quality speech recordings were collected in a studio environment. Several audio effects were further applied to these recordings by professional sound engineers to make them sound more aesthetically pleasing.} and a number of versions of low-quality speech\footnote{Low-quality speech recordings were produced by replaying high-quality audio through a professional loudspeaker and re-recording them with different consumer devices in different real-world acoustic environments.}, which are affected by noise, reverberation, and microphone response. Specifically, it consists of 20 speakers (10 female and 10 male) reading 5 excerpts each from public domain stories. 
To prepare the training set, we selected 4 of the 5 excerpts narrated by 18 of the 20 speakers under 7 of the 10 recording conditions then split the corresponding recordings into shorter segments, which resulted in 23,555 audio clips. The remaining 1 excerpt, 2 speakers (1 female and 1 male), and 3 conditions were used to form the test set, which resulted in 228 audio clips. Thus, all the content, speakers, and recording conditions of the tested speech were unseen to the training set. The three tested real-world recording conditions were: (1) \textit{ipad\_livingroom}, recording was done by an iPad Air in a living room; (2) \textit{ipadflat\_office}, recording was done by an iPad Air placed flat in an office; and (3) \textit{iphone\_bedroom}, recording was done by an iPhone 5S in a bedroom. 
\vspace{-2.0mm}
\subsection{Implementation}
\label{sec:implement}
\vspace{-0.5mm}
All audios were resampled at 16kHz. We used STFT to compute the spectrogram with a Hanning window size of 50 ms and a hop size of 12$.$5 ms, and the spectrogram was power-law compressed \cite{wilson2018exploring} with a power of 0.3. For WaveRNN vocoder, we used a public speaker-independent model\footnote{\url{https://github.com/erogol/WaveRNN}}, which was pretrained sufficiently with more than 900 speakers selected from the LibriTTS corpus \cite{zen2019libritts}. We slightly fine-tuned the model, with the high-quality studio recordings in the training set, to make the model adapt to the studio audio effect. Note that the speakers and content of the tested recordings were still unseen to the fine-tuned \mbox{WaveRNN} vocoder.

Although the primary target of this work is to enhance low-quality recordings, we implemented audio effect transfer, e.g., transferring the iPhone recording in the bedroom to the iPad recording in the office, within one unified system. As shown in Fig.~\ref{fig:diagram}, the decoder can predict not only the Mel spectrogram in studio quality but also that under other recording conditions, depending on the reference audio\footnote{The reference audio was randomly selected, which did not correlate with the input recording in terms of both speaker and content.}. 
This architecture enables us to augment training data with diverse combinations of input and reference pairs. Since the system learns to disentangle the channel factor and adapt to various recording conditions, we expect that it can reduce overfitting and improve overall performance. 
Therefore, each audio clip under 7 training recording conditions was combined with 3 different types of references: one high-quality recording (as primary training target) and two recordings that were randomly selected from the other 6 conditions. This extended the original training set and resulted in a total of 70,665 (23,555 $\times$ 3) training examples. For the test set, we set the high-quality recording only as the reference since our ultimate target is to examine if the low-quality input can be enhanced. 
The Adam optimizer \cite{kingma2014adam} was used for training, with learning rates of $0.0001$ and $0.0002$ for the model and its classifiers, respectively. Hyper-parameters $\alpha$, $\beta$, and $\gamma$ in Eq.\ \ref{eq:all-obj} were set to $1.0$, $0.2$, and $0.05$, respectively. 

\begin{table}[t]
\vspace{-1.8mm}
    \caption{Objective evaluation results of different systems on test set. For all four measures, higher scores indicate better performance.}
    \vspace{1mm}
    \label{tab:oe}
    \centering
    \renewcommand\arraystretch{1.25}
    \scalebox{1.03}{
    \begin{tabular}{l c c c r}
        \toprule
          \textbf{System} &
           \textbf{CSIG} &
           \textbf{CBAK} &
           \textbf{COVL} & \textbf{STOI} \\
           \midrule
           \textbf{Raw audio} & 3.05 & 2.23 & 2.60 & 0.869 \\
           \midrule
           \textbf{WPE} & 3.16 & 2.41 & 2.75 & 0.888 \\
           \textbf{WPE+L} & 2.81 & 2.33 & 2.52 & 0.811 \\
           \midrule
           \textbf{Wavenet} & 3.67 & 2.42 & 3.08 & 0.904 \\
           \midrule
           \textbf{Linear-ISTFT} & 3.94 & 2.61 & 3.37 & 0.905 \\
           \midrule
           \textbf{ED} & 3.89 & 2.48 & 3.28 & 0.906 \\
           \textbf{ED+CM} & 3.73 & 2.49 & 3.16 & 0.886 \\
           \textbf{FULL} & 3.94 & 2.52 & 3.34 & 0.906 \\
        \bottomrule
    \end{tabular}
    }
    \vspace{-2mm}
\end{table}

\vspace{-2.0mm}
\subsection{Evaluated Systems}
\label{sec:es}
\vspace{-0.5mm}
We conducted an ablation study on the proposed system. Several speech enhancement baseline systems were also reimplemented, making a total of seven systems compared in our experiments. We describe and notate each system as follows:
\vspace{-6.0mm}
\begin{itemize}
    \item \textbf{ED}: A simplified version of our proposed system that is composed only of encoder and decoder modules. The decoder only predicts the high-quality Mel spectrogram as its prediction target.
    \vspace{-1.25mm}
    \item \textbf{ED+CM}: Another simplified version that is composed of encoder, decoder, and CM modules. No classifiers and corresponding training objectives were used for training. Following the work of \cite{su2020acoustic}, we improved this system by conditioning the encoder with the input's channel factor\footnote{As an alternative to adversarial training, this additional conditioning was used to encourage the encoder to produce channel-invariant features.}.
    \vspace{-1.25mm}
    \item \textbf{FULL (ED+CM+Classifiers)}: Our complete proposed system shown in Fig.~\ref{fig:diagram}, which consists of an encoder, decoder, CM network, and three classifiers. Auxiliary (Eq.\ \ref{eq:eq6}) and adversarial (Eq.\ \ref{eq:eq5} and Eq.\ \ref{eq:eq7}) training objectives were integrated through these three classifiers.
    \vspace{-5.5mm}
    \item \textbf{Linear-ISTFT}: This system shares the same settings as \texttt{FULL}, except the decoder output was changed to linear spectrogram. Instead of WaveRNN, we synthesized the waveform using ISTFT with the noisy phase.
    \vspace{-1.25mm}
    \item \textbf{Wavenet}: A waveform-to-waveform mapping system based on Wavenet \cite{su2019perceptually}. We reimplemented it with the same model architecture and training objective (L1 loss on log spectrogram).
    \vspace{-1.25mm}
    \item \textbf{WPE}: A state-of-the-art speech de-reverberation baseline, which estimates a linear filter to minimize the weighted linear prediction error \cite{nakatani2010speech}.
    \vspace{-1.25mm}
    \item \textbf{WPE+L}: An integrated system that sequentially combines \texttt{WPE} for de-reverberation and a standard log-MMSE speech magnitude estimator \cite{ephraim1985speech} for denoising.
    \vspace{-1.25mm}
\end{itemize}

\vspace{-5mm}
\subsection{Objective Evaluations}
\label{sec:oe}
\vspace{-0.5mm}
We first evaluated each system with objective measures. We used the short-time objective intelligibility (STOI) score \cite{taal2010short} to measure speech intelligibility and three composite scores (CSIG, CBAK, and COVL) \cite{hu2006evaluation} to measure enhancement quality. CSIG, CBAK, and COVL are mean opinion score (MOS) predictions of speech distortion, noise distortion, and overall quality, respectively. The evaluation results are listed in Table~\ref{tab:oe}.

As shown, the \texttt{FULL} system consistently improves its two simplified versions (\texttt{ED} and \texttt{ED+CM}) for all measures, which indicates both the CM network and classifiers play important roles in our proposed system. It also significantly outperforms time-domain \texttt{Wavenet} and the two signal-processing baselines (\texttt{WPE} and \texttt{WPE+L}). \texttt{WPE+L} system performs much worse than \texttt{WPE}. This is mostly because the log-MMSE estimator suppresses noise too aggressively even though the noise level of the DAPS dataset is not high, therefore it degrades speech quality. We found that \texttt{FULL} system is worse than \texttt{Linear-ISTFT} in terms of CBAK and COVL. The probable reason is that the vocoder-generated speech has more artifacts than the ISTFT-generated one. However, most of these artifacts introduced by the neural vocoder do not affect human perception, as has been observed in our previous work \cite{li2020noise}. 
To comprehensively evaluate each system, we further conducted the following subjective listening tests.

\vspace{-2.5mm}
\subsection{Subjective Evaluations}
\label{sec:se}
\vspace{-0.5mm}
\begin{figure}[t]
\begin{minipage}[b]{0.99\linewidth}
  \centering
  \centerline{\includegraphics[width=0.99\linewidth]{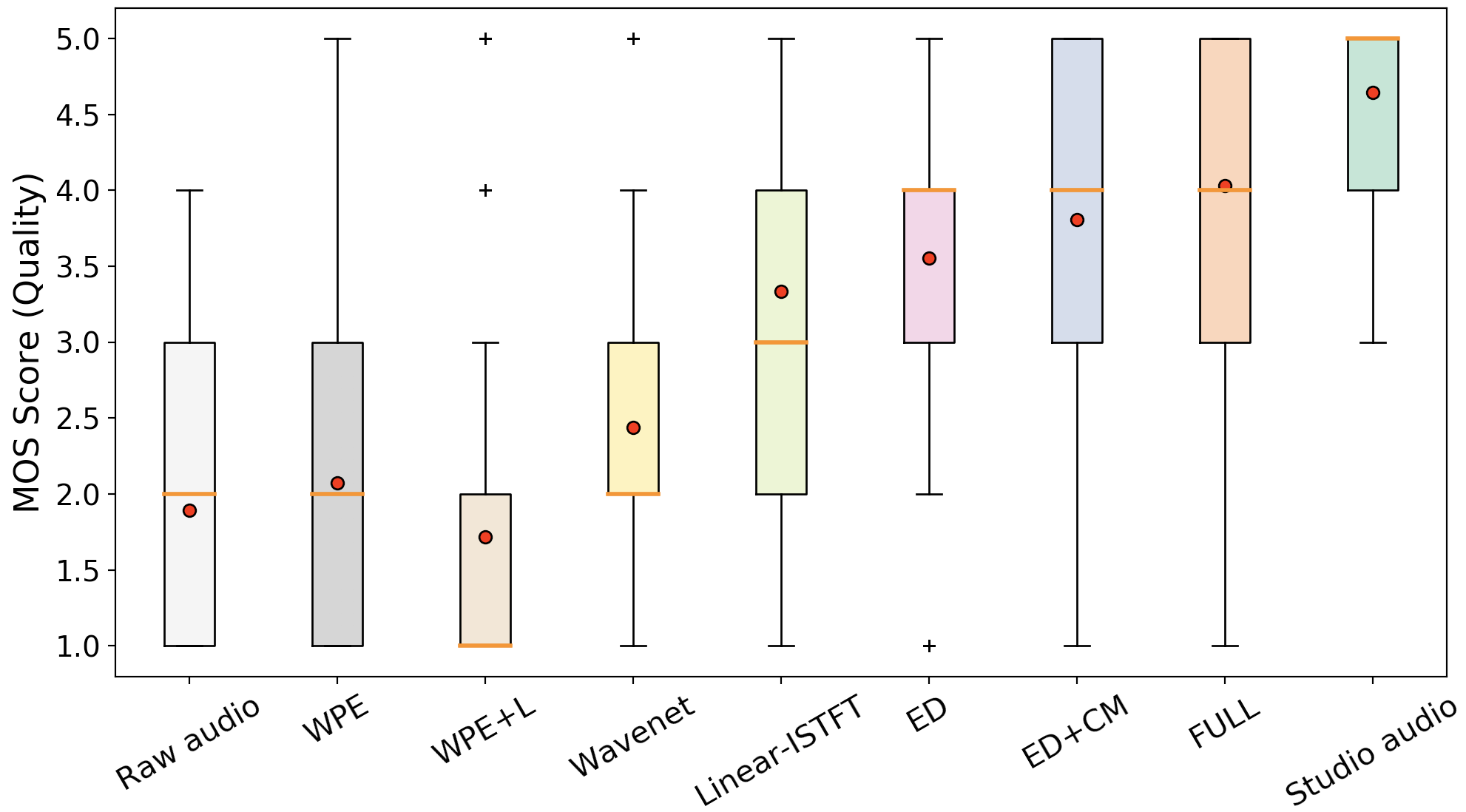}}
\end{minipage}
\caption{Box plot on MOS scores for audio quality. Red dots represent mean score.}
\label{fig:mos}
\vspace{-3mm}
\end{figure}
We conducted crowdsourced listening tests for the subjective evaluations. Specifically, we chose 120 (20 audios $\times$ 3 conditions $\times$ 2 genders) of the 228 tested audio clips for each system\footnote{Audio samples are available at: \url{https://nii-yamagishilab.github.io/hyli666-demos/evr-slt2021}}. Participants (165 individuals) were asked to rate the quality of each anonymized audio from 1-5 (five-point Likert scale) for the mean opinion score (MOS). For reference, the raw (with low quality) and studio versions of each audio were also provided to the participants before rating. Each audio was rated ten times to avoid human bias. 

The listening results are shown in Fig.~\ref{fig:mos}. The Mann-Whitney U test \cite{nachar2008mann} reveals that the proposed \texttt{FULL} system significantly outperforms the other systems with \textit{p}-values all lower than 1e-7. 
It is noteworthy that unlike the objective results, \texttt{FULL} system shows a higher score than \texttt{Linear-ISTFT}, which means the \mbox{WaveRNN} module successfully improves the quality of the synthetic waveform. This also indicates that although the artifacts introduced by \mbox{WaveRNN} degrade the objective results, they do not affect human subjective evaluations.
More interestingly, we can see that the \texttt{FULL} system outperforms \texttt{ED} in both subjective and objective tests, even though the task of \texttt{FULL} system is more challenging: the extra channel factor should be disentangled, and the output Mel spectrogram can be not only in studio quality but also in other acoustic characteristics based on the provided reference. Such additional learned knowledge related to channel information does benefit the \texttt{FULL} system and improves its performance.

\vspace{-2mm}
\subsection{Beyond Enhancement: Audio Effect Transfer}
\label{sec:aet}
In addition to speech enhancement, the proposed system can also realize audio effect transfer: transferring the input recordings to sound as if they were recorded in another environment. To achieve this, we only need a few or even one reference audio recorded under the corresponding desired channel (recording) condition. 

\begin{figure}[t]
\begin{minipage}[b]{0.99\linewidth}
  \centering
  \centerline{\includegraphics[width=0.82\linewidth]{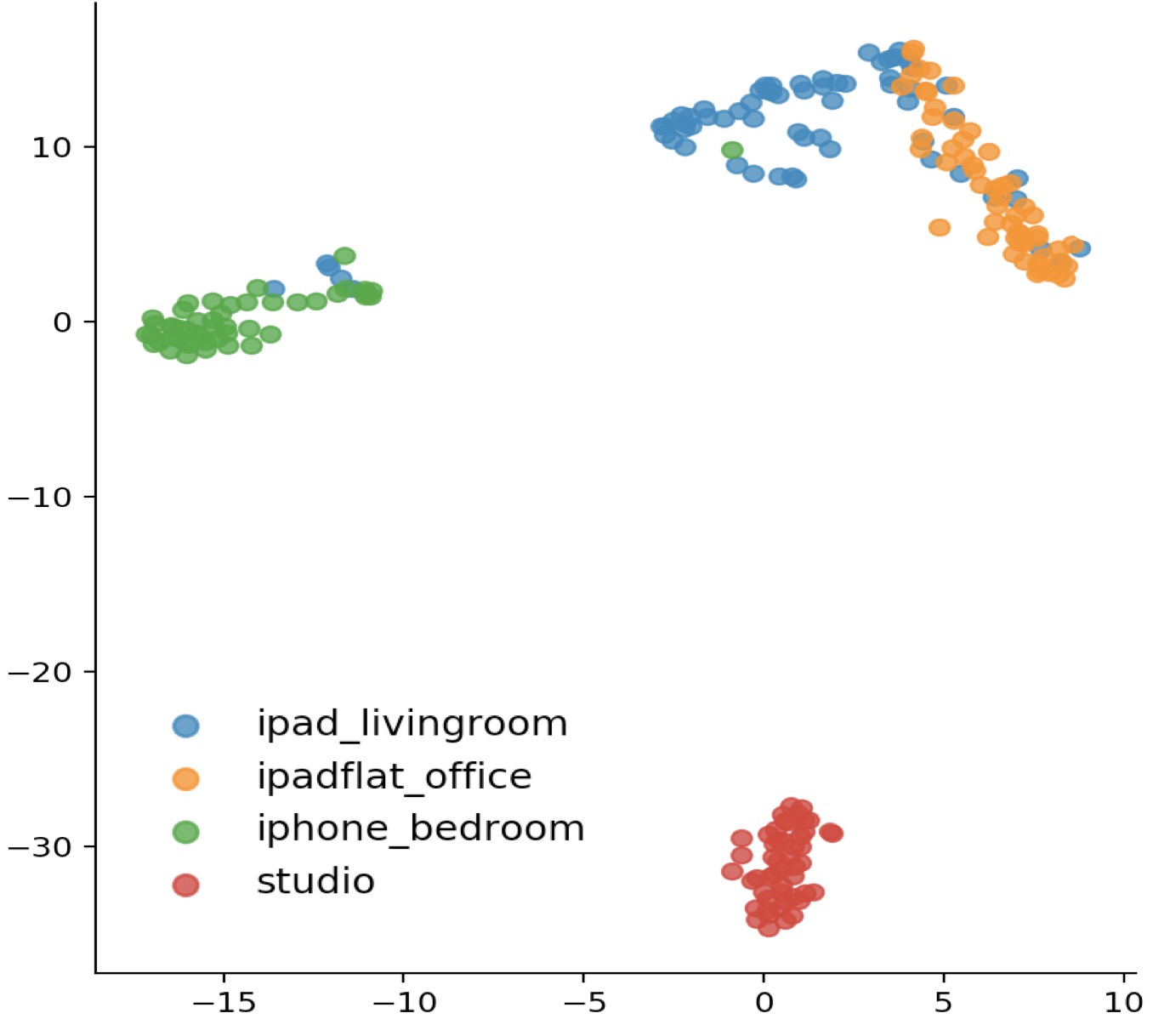}}
\end{minipage}
\caption{Visualization of learned channel factors using t-SNE transformation; color coded by channel condition labels.}
\label{fig:tSNE}
\vspace{-2mm}
\end{figure}

\begin{figure}[t]
\begin{minipage}[b]{.496\linewidth}
  \centering
  \centerline{\includegraphics[width=\linewidth]{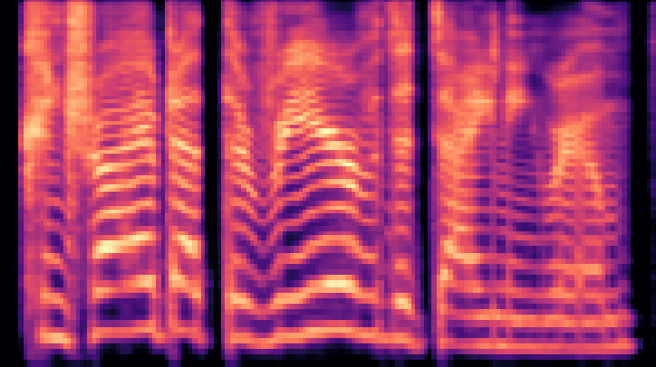}}
  \centerline{(a) Studio}\medskip
\end{minipage}
\hfill
\begin{minipage}[b]{.496\linewidth}
  \centering
  \centerline{\includegraphics[width=\linewidth]{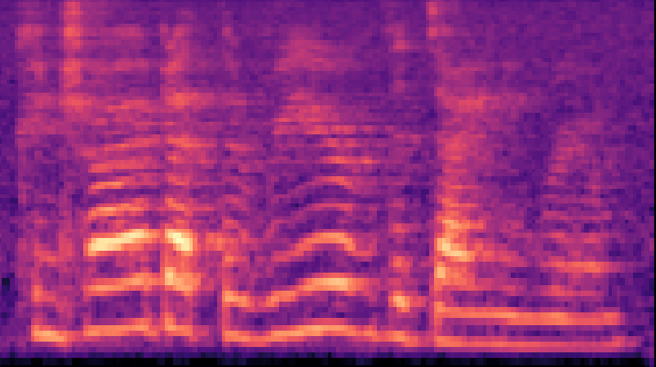}}
  \centerline{(b) Transfer target}\medskip
\end{minipage}
\begin{minipage}[b]{.496\linewidth}
  \centering
  \centerline{\includegraphics[width=\linewidth]{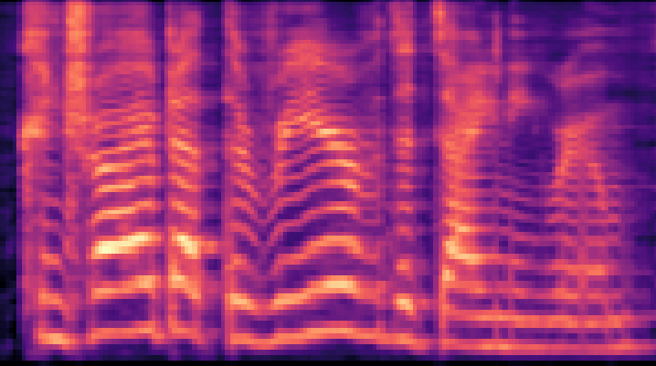}}
  \centerline{(c) Transferred at $\alpha=0.6$}\medskip
\end{minipage}
\begin{minipage}[b]{0.496\linewidth}
  \centering
  \centerline{\includegraphics[width=\linewidth]{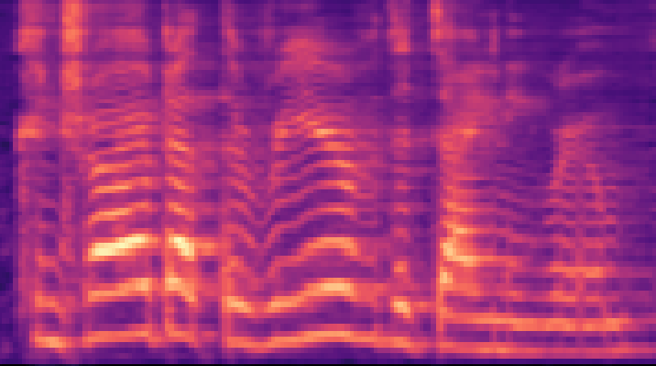}}
  \centerline{(d) Transferred at $\alpha=1.0$}\medskip
\end{minipage}
\vspace{-5.0mm}
\caption{Examples of transferred Mel spectrograms: (a) Studio recording, (b) Target recording recorded in \textit{iphone\_bedroom} condition, (c) Transferred recording with $\alpha=0.6$, and (d) Transferred recording with $\alpha=1.0$.}
\label{fig:transferred-specs}
\vspace{-3mm}
\end{figure}

Instead of using one-hot code, the CM network automatically encodes the channel factor from the arbitrary reference, then the decoder can predict the target-environment Mel spectrogram conditioned on this factor. Figure~\ref{fig:tSNE} gives a visualization of learned channel factors for different reference recordings under three tested unseen conditions and one studio condition. We used t-SNE transformation \cite{maaten2008visualizing} to project the 256-dimensional channel factor into 2 dimensions. We can see that the learned factors are clearly clustered based on their channel conditions\footnote{There is a little overlap between the conditions of \textit{ipad\_livingroom} and \textit{ipadflat\_office}. This is because recording device used under both conditions was same (iPad Air), which leads to relatively similar channel characteristics.}, which indicates that the CM network can effectively discriminate unseen reference audios and produce representative factors. Therefore, it enables the system to deal with the unlabelled references under unseen channel conditions. With this system, we can further control the transferred effect (e.g. reverberation level) by flexibly scaling the channel factor. Examples of transferred Mel spectrograms are given in Fig.~\ref{fig:transferred-specs}, where we aimed to transfer a studio recording to sound as if it were recorded in the (unseen) \textit{iphone\_bedroom} condition. Instead of feeding a reference audio, the applied channel factor $\boldsymbol{\hat{z}}_c$ was pre-computed through linear interpolation of two factors using Eq.\ \ref{eq:interpolation}.
\begin{equation}
    \boldsymbol{\hat{z}}_c = (1-\alpha) * \boldsymbol{z}_c^{pro} + \alpha * \boldsymbol{z}_c^{iph}
\label{eq:interpolation}
\end{equation}
where $\boldsymbol{z}_{c}^{pro}$ and $\boldsymbol{z}_c^{iph}$ denote the channel factors extracted from a professional studio recording and \textit{iphone\_bedroom} recording, respectively, and $\alpha$ is the scale value that ranges from $0$ to $1$. We successfully controlled the transferred effect from less reverberant (Fig.~\ref{fig:transferred-specs}~(c) ) to more reverberant (Fig.~\ref{fig:transferred-specs}~(d)) by increasing the scale value $\alpha$. We can also see that the transferred Mel spectrogram in Fig.~\ref{fig:transferred-specs}~(d) shares a similar audio effect (or channel characteristics) with the ground-truth transfer target in Fig.~\ref{fig:transferred-specs}~(b). 

\vspace{-2mm}
\section{Conclusions}
\label{sec:conclu}
\vspace{-0.5mm}
In this paper, we proposed a system to enhance low-quality voice recordings. Specifically, the channel factor disentangled from a high-quality reference recording is used to guide the system to predict the enhanced Mel spectrogram, which is then transformed to the enhanced waveform via a \mbox{WaveRNN} vocoder. Experimental results show that our system works well and outperforms several state-of-the-art baselines. Moreover, we show that it can be flexibly extended to transform the input recording into not only professional studio quality (as our primary target) but also with other acoustic (or channel) characteristics based on the reference we designate. 

Our future work includes improving the naturalness of the predicted Mel spectrogram by adopting a generative adversarial network-based spectrogram discriminator \cite{su2019perceptually, su2020hifi}. 
In addition, we found that the synthetic waveform has lower perceived quality if it is synthesized from the reverberant spectrogram. This is because our neural vocoder was pretrained only with high-quality but dry waveforms. We plan to alleviate this issue by using a recently proposed reverberation-aware vocoder \cite{ai2020reverberation}, which is able to model reverberation and generate a reverberant waveform with high perceived quality.

\vspace{1mm}
\noindent
\textbf{Acknowledgments}
\sloppy
This work was partially supported by a JST CREST Grant (JPMJCR18A6, VoicePersonae project), Japan, and MEXT KAKENHI Grants (16H06302, 17H04687, 18H04120, 18H04112, 18KT0051, 19K24372), Japan. The experiments were partially conducted on TSUBAME 3.0 of Tokyo Institute of Technology.

\vfill\pagebreak

\bibliography{main.bbl}

\end{document}